\documentclass[aps,prl,twocolumn,groupedaddress,superscriptaddress]{revtex4}
\usepackage{graphicx}
\usepackage[colorlinks]{hyperref}
\usepackage{amssymb}
\usepackage{ctable}
\usepackage{amsmath}
\usepackage{amsfonts}
\usepackage{amssymb}
\usepackage{multirow}
\usepackage{makecell}

\begin{document}

\title{Hole-lattice Coupling and Photo-induced Insulator-Metal Transition in VO$_2$}

\author{Xun Yuan}
\affiliation{State Key Laboratory of High Performance Ceramics and Superfine Microstructures, 
Shanghai Institute of Ceramics, Chinese Academy of Sciences, Shanghai 200050, China}

\author{Wenqing Zhang}
%\email{wqzhang@mail.sic.ac.cn}

\affiliation{School of Chemistry $\&$ Chemical Engineering, and Sate Key Laboratory of Coordination Chemistry,
Nanjing University, Jiangsu 210093, China}
\affiliation{State Key Laboratory of High Performance Ceramics and Superfine Microstructures, 
Shanghai Institute of Ceramics, Chinese Academy of Sciences, Shanghai 200050, China}

\author{Peihong Zhang}
%\email{pzhang3@buffalo.edu}

\affiliation{Department of Physics, University at Buffalo, State University of New York, Buffalo, New York 14260, USA}

\date{\today}

\begin{abstract}
Photo-induced insulator-metal transition in VO$_2$ and the related 
transient and multi-timescale structural dynamics upon photoexcitation 
are explained within a unified framework. Holes created by photoexcitation 
weaken the V-V bonds and eventually break V-V dimers in the M$_1$ phase of
VO$_2$ when the laser fluence reaches a critical value.
The breaking of the V-V bonds in turn leads to an immediate
electronic phase transition from an insulating to a metallic state while
the crystal lattice remains monoclinic in shape. The coupling between
excited electrons and the 6.0 THz phonon mode is found to be responsible 
for the observed zig-zag motion of V atoms upon photoexcitation and is consistent with
coherent phonon experiments.
\end{abstract}

\maketitle

%\section{Introduction}

Despite decades of intensive research, the physics behind the metal-insulator transition in VO$_2$ at about 
340 K\cite{PhysRevLett.3.34} remains a subject of unabated debate. The electronic phase transition is 
accompanied by a seemingly ``simultaneous" structural change from a low temperature monoclinic
(M$_1$) to a high temperature rutile (R) structure. It is likely that the phase transition involves
multiple intermediate states that occur at distinct timescales. 
Unfortunately, the delicate and transient interplay between
the atomic and electronic degrees of freedom, which determines the dynamics and the kinetic
pathway of the phase transition in VO$_2$, cannot be easily accessed within time averaged or equilibrium 
measurements. Therefore, recent advances in ultrafast spectroscopy have brought much excitement and inspired a new wave of 
investigations~\cite{PhysRevLett.87.237401,PhysRevB.69.153106,PhysRevB.70.161102, PhysRevLett.99.116401, PhysRevB.76.035104,
nakajima011907,PhysRevB.83.195120,PhysRevB.85.155120,Baum02112007,PhysRevLett.97.266401,PhysRevLett.109.166406}
with unprecedented time resolution and accuracy. 

These new experiments reveal a great deal of details on the ultrafast dynamics and intermediate states
associated with the phase transition, offering rare insights into the intriguing physics of VO$_2$. 
For example, an ultrafast ($\sim 100$ fs) photo-induced 
insulator-metal transition was observed~\cite{PhysRevLett.87.237401,PhysRevB.69.153106,PhysRevB.70.161102, 
PhysRevLett.99.116401, PhysRevB.76.035104, nakajima011907,PhysRevB.83.195120,PhysRevB.85.155120}. The observed response 
cannot be explained by simply considering the effects of the excited carriers, and a structurally driven 
phase transition mechanism was proposed~\cite{PhysRevB.70.161102}. The structural change was ascribed to coherently 
generated optical phonons at ultrashort time scales.
Indeed, a coherent phonon mode with frequencies at about 6.0 THz ($\sim$ 200 cm$^{-1}$) 
is often observed~\cite{PhysRevB.70.161102,PhysRevB.83.195120,PhysRevLett.97.266401,PhysRevLett.99.116401}.
Other coherent phonons at about 4.5 THz ($\sim$ 150 cm$^{-1}$)~\cite{PhysRevLett.97.266401}
and 6.75 THz ($\sim$ 225 cm$^{-1}$)~\cite{PhysRevB.70.161102} have also been observed.
However, the precise role these coherently generated phonons play and the mechanism of their generation upon photoexcitation
are still not well understood. 

%\begin{table}
%\setlength{\tabcolsep}{2.7pt}
%\caption{\label{tab:M1struc}Relaxed crystal structure of the M$_1$ phase
%(space group: $P2_1/C$). The lattice constants ($a$, $b$, $c$) are in units
%of $\mathrm{\AA}$. The experimental structure (shown in parentheses) 
%is taken from Ref.\cite{M1struc}.}
%\begin{tabular}{cccc}
%\hline\hline
%$\beta$&$a$ & $b$ & $c$  \\
%\hline
%122.01$^\circ$ & 5.653   & 4.596   & 5.421\\
%(122.65$^\circ$) &( 5.752) & (4.538) & (5.383)\\
%\hline
%Atom  & $x$ & $y$ & $z$ \\
%\hline
%~V     & 0.2339 (0.2395) & 0.9748 (0.9789) & 0.0285 (0.0265) \\
%O$_1$ & 0.1121 (0.1062) & 0.2158 (0.2119) & 0.2114 (0.2086) \\
%O$_2$ & 0.4022 (0.4005) & 0.7011 (0.7026) & 0.2969 (0.2988) \\
%\hline\hline
%\end{tabular}
%\end{table}
 
More recently, a 4D visualization of the transitional structures of VO$_2$ after 
photoexcitation was carried out using a femtosecond electron diffraction technique~\cite{Baum02112007}. 
A multi-timescale structural evolution ranging from subpicosecond to nanoseconds was uncovered.
The first step of the structural change shortly after photoexcitation
was identified as a rapid (subpicosecond) separation of the initially paired
V atoms (i.e., V-V dimers) in the M$_1$ phase VO$_2$. It is natural to associate 
this subpicosecond displacements of V atoms with the observed
ultrafast ($\sim 100$ fs) photo-induced insulator-metal transition.
Unfortunately, there is no direct evidence that supports this connection.

In this letter, results from first-principles electronic structure calculations are used to
establish a theory that is able to explain the ultrafast 
photo-induced insulator-metal transition and the the multi-timescale structural dynamics 
associated with photoexcitations. The strong coupling between the lattice and
the excited holes is responsible for the observed rapid separation of V-V pairs after photoexcitation.
The atomic motion is found to be primarily associated with a 339 cm$^{-1}$ (10.2 THz) phonon mode,
instead of the much discussed 6.0 THz ($\sim$ 200 cm$^{-1}$) mode.
The separation of the V-V pair in-turn leads to an immediate electronic phase transition.
 although the crystal lattice remains monoclinic in shape.
The longer timescale ($\sim$ a few picoseconds) structural dynamics observed in experiment 
is associated with a phonon mode at 200 cm$^{-1}$ ($\sim$ 6.0 THz) often observed in coherent phonon 
experiments~\cite{PhysRevB.70.161102,PhysRevB.83.195120,PhysRevLett.97.266401,PhysRevLett.99.116401}.
This phonon mode involves zig-zag movements of V atoms and is found to couple primarily with 
the excited electrons. However, the excitation of this
phonon mode is not responsible for the subpicosecond photo-induced insulator-metal transition.
Instead, it is just an important step toward the full structural phase transition 
from the monoclinic to the rutile phase.

\begin{figure}
\includegraphics[width=7cm]{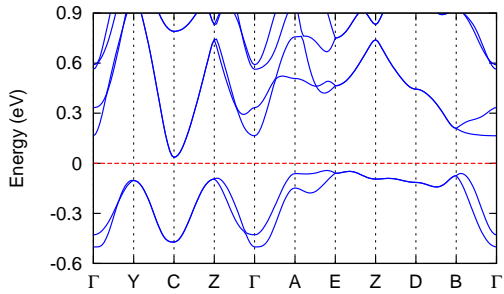}
\caption{\label{fig:M1es} (Color online) 
Calculated band structure of the M$_1$ phase VO$_2$ using theoretically relaxed crystal structure.
}
\end{figure}

All calculations are carried out using the projector augmented wave method~\cite{PhysRevB.50.17953}
implemented in VASP~\cite{PhysRevB.54.11169}, and the Perdew-Burke-Ernzerhof (PBE) 
functional is used. It is worth mentioning that although the PBE functional is not
able to produce an insulating ground state for the M$_1$ phase VO$_2$ if the experimental
structure is used, upon structural relaxation, 
a small band gap ($\sim 0.13$ eV) actually develops, as shown in Fig~\ref{fig:M1es}.
The use of the theoretically relaxed structure also gives a satisfactory description of the lattice 
dynamics of the M$_1$ phase VO$_2$. Table~\ref{tab:M1G} compares the calculated zone-center phonon 
frequencies with experiment. The agreement between theory and experiment is very good.

Photoexcitations naturally introduce hot electrons and holes into the system. However, as 
mentioned earlier, the observed optical response upon photoexcitation cannot be explained by simply
considering the effects of these excited electrons/holes~\cite{PhysRevB.70.161102}, and 
the transient lattice dynamics must be taken into account.
In order to separate the effects of the excited electrons and holes on the 
transient lattice dynamics of the system, in particular, the observed rapid separation of V-V pairs, 
we introduce electrons and hole into the system in our calculation separately. 
Our calculations start with a fully relaxed M$_1$ structure. Additional electrons or holes
are then introduced into the system, and the structures are fully relaxed
for a given electron (or hole) ``doping" level while keeping the monoclinic lattice vectors. 

\begin{table}[b!]
\setlength{\tabcolsep}{3.7pt}
\caption{\label{tab:M1G} Calculated zone-center phonon modes of the M$_1$ phase VO$_2$. 
Zone-center phonons are grouped into Raman active (A$_g$ or B$_g$) 
and infrared active (A$_u$ or B$_u$) modes.
Experimental results are taken from Ref.\cite{Schilbe2004449} (Raman)
and Ref.\cite{PhysRevLett.17.1286} (infrared). The projections $\eta$
of the A$_g$ modes are discussed in the text.}
\begin{tabular}{ccccccccccccc}
  \hline\hline
  \multicolumn{3}{c}{A$_g$} && \multicolumn{2}{c}{B$_g$} && \multicolumn{2}{c}{A$_u$} && \multicolumn{2}{c}{B$_u$} \\
  \cline{1-3}\cline{5-6}\cline{8-9}\cline{11-12}
  Cal. & Exp.& $\eta$&& Cal. & Exp. && Cal. & Exp. && Cal. & Exp.  \\
  \hline 
  152 & 149&0.54&& 212 & $...$ && $...$&  $...$ && $...$ &  $...$  \\
  197 & 199&2.72&& 231 & 259 && 187&  189 && $...$ &  $...$  \\
  224 & 225&0.96&& 247 & 265 && 265&  270 && 253 &  227  \\
  331 & 313&0.59&& 374 & 395 && 304&  310 && 295 &  285  \\
  339 & 339&1.58&& 432 & 444 && 333&  340 && 324 &  324  \\
  389 & 392&0.82&& 447 & 453 && 415&  392 && 374 &  355  \\
  508 & 503&0.79&& 495 & 489 && 508&  505 && 490 &  478  \\
  605 & 618&0.26&& 593 & 595 && 560&  600 && 572 &  530  \\
  675 & 670&0.11&& 758 & 830 && 721&  710 && 721 &  700  \\
  \hline\hline
  \end{tabular}
\end{table}

\begin{figure}[!b]
\includegraphics[width=7cm]{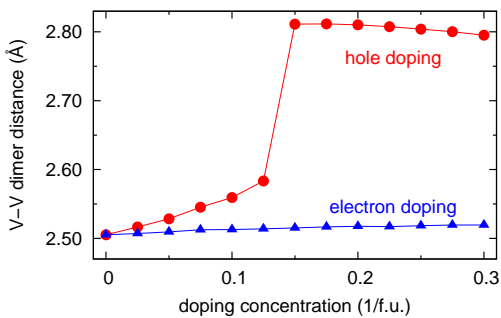}
\caption{\label{fig:disdop} (Color online) 
Relaxed V-V dimer distance after introducing holes or electrons to the system.}
\end{figure}

Figure~\ref{fig:disdop} illustrates the drastically 
different effects of electrons and holes on
the calculated V-V separation: Whereas the V-V separation is nearly unchanged upon
electron doping, hole doping results in a gradual elongation of the V-V bond length
below a critical level of $\sim$ 0.15 holes per VO$_2$ formula unit, above which a 
sudden jump in V-V separation is observed. The existence of a critical hole level 
and the sudden jump in V-V separation may be related to the required critical excitation 
fluence in experiments~\cite{Baum02112007,PhysRevLett.99.116401}.
Therefore, we conclude that the observed rapid V-V separation upon photoexcitation
is a result of a strong lattice-hole coupling in this system. 

\begin{figure*}
\includegraphics[width=14cm]{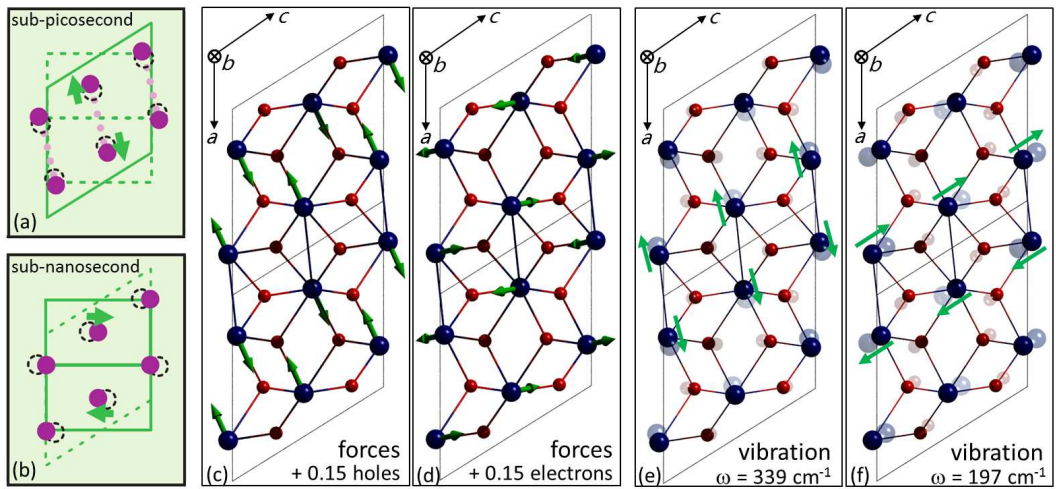}
\caption{\label{fig:VIB} (Color online) Atomic displacements upon photoexcitation: comparison between
theory and experiment. Left panels: the observed subpicosecond (a) and subnanosecond (b) 
movement of V atoms in the M$_1$ phase VO$_2$ upon photoexcitation~\cite{Baum02112007}.
Middle panels: calculated initial forces on V atoms with additional holes (c) and electrons (d).
Right panels: atomic displacement (schematic) of the 339 cm$^{-1}$ (e) and the 197 cm$^{-1}$ phonon modes (f).
The directions of the displacements of V atoms are shown for comparison. 
The V and O atoms are shown with large (dark blue) and small (red) balls, respectively. 
Atoms after displacement are shown with light colors.}
\end{figure*}

In order to relate our results to the observed multi-timescale atomic
motion after photoexcitations~\cite{Baum02112007}, 
we compare the experimentally observed atomic displacements [Figs.~\ref{fig:VIB}(a) and \ref{fig:VIB}(b)]
at two distinct timescales and the directions of the calculated initial forces [Figs.~\ref{fig:VIB}(c) and \ref{fig:VIB}(d)] on
atoms with electron or hole doping with a ``doping" level (carrier density) of
0.15 per VO$_2$ formula unit. The calculation starts with the relaxed M$_1$ structure 
before introducing carriers into the system. Carriers are then introduced, 
the forces on atoms are then calculated. Upon hole doping, the calculated forces on V atoms
are directed primarily along the direction of V-V pairs 
[the two V atoms in the middle Fig.~\ref{fig:VIB}(c)] and they tend to elongate 
the V-V bonds. This result coincides nicely with the observed~\cite{Baum02112007} 
subpicosecond displacement of V atoms upon photoexcitation of VO$_2$ as shown in Fig.~\ref{fig:VIB}(a).
Electron doping, on the other hand, results in forces on V atoms that are nearly
parallel to the lattice $c$ direction and zig-zag along the lattice $a$ direction, as shown
in Fig.~\ref{fig:VIB}(d). This result compares well with the observed
subnanosecond atomic displacements as shown in Fig.~\ref{fig:VIB}(b).
Therefore, our results clearly illustrate that the subpicosecond V-V separation comes from
the coupling between lattice and holes, whereas the subnanosecond
structural dynamics should be attributed to the electron-lattice coupling and is related to the
zig-zag motion of the V atoms. 

\begin{figure}[b!]
\includegraphics[width=7cm]{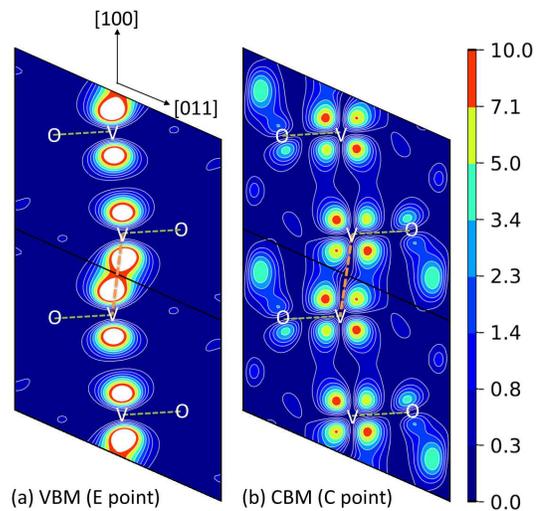}
\caption{\label{fig:VBMCBM} (Color online) Charge density of the 
VBM (a) and the CBM (b) states on the (-110) plane of the M$_1$ phase.
The shortest V-O and V-V bonds are approximately 
on the (-110) plane and are shown with dashed lines.}
\end{figure}

This disparate structural responses to electrons and holes can be
understood by investigating the bonding character of the valence band maximum
(VBM) and the conduction band minimum (CBM) states as shown in Fig.~\ref{fig:VBMCBM}.
The VBM states are primarily of V-V bonding character [Fig.~\ref{fig:VBMCBM}(a)].
Therefore, removing electrons from the VBM states (adding holes) results 
in a weakening of V-V bonds. When this weakening
reaches a critical point, the V-V dimerization is no longer intact.
This explains the sudden jump in the V-V separation when hole level
reaches a critical value of about 0.15 holes per VO$_2$ formula unit as shown in Fig.~\ref{fig:disdop}.
This critical hole density may also be related to the critical lase fluence in experiment.
The CBM states, on the other hand, involve an antibonding character of
the shortest V-O bonds, as shown in Fig.~\ref{fig:VBMCBM}(b). These short V-O bonds are related to
the zig-zag arrangement of the V atoms along the lattice $a$ direction
as shown in Fig.~\ref{fig:VIB}. Therefore, adding  electrons to the antibonding V-O 
states tends to alleviate the zig-zag arrangement of the V atoms. These results
are consistent with the force analysis shown in Figs.~\ref{fig:VIB}(c) and \ref{fig:VIB}(d); they also agree
with the measured atomic displacement as shown in Figs.~\ref{fig:VIB}(a) and \ref{fig:VIB}(d).  

Although the transient structural dynamics involved in the photo-induced insulator-metal
transition of VO$_2$ is very different from the conventional description of lattice dynamics 
in terms of phonons, it is still instructive to map the structural change onto 
polarization vectors of relevant phonon modes. Here we define a displacement vector
from the M$_1$ to the R phase: 
\begin{equation}
\Delta {\mathbf{R}}\equiv \{\vec{R_i}^{\mathrm{R}}-\vec{R_i}^{\mathrm{M}_1}\}; ~i=1,2,...,N,
\label{eq:eq1}
\end{equation}
where $\vec{R_i}$ is the position of the $i$th atom in the unit cell and 
$N$ is total number of atoms. We double the size of the unit cell of the R phase so that 
there is one-to-one correspondence between atoms in both phases.
As it was suggested experimentally~\cite{Baum02112007}, 
changes in internal structure after photoexcitations happen much faster than 
the change in the shape of the lattice. Therefore, in order to analyze the
initial structural evolution, we use the lattice vectors of
the M$_1$ phase. Symmetry requires that the displacement vector
$\Delta {\mathbf{R}}$ has the A$_g$ symmetry. Therefore, this displacement vector
can only be projected onto Raman active A$_g$ phonons. The overlap between a
Raman active phonon $i$ and the displacement vector $\Delta {\mathbf{R}}$ is calculated via
\begin{equation}
\eta_i=\sum_j \frac{1}{\sqrt{m_j}}\vec{e}_{ij}\cdot\Delta{\mathbf{R}},
\label{eq:eq2}
\end{equation}
where $m_j$ is the mass of the $j$th atom, $\vec{e}_{ij}$ is $j$th atomic 
component of the polarization vector of the $i$th phonon mode.
The result of this projection of the atomic displacement is shown in Table~\ref{tab:M1G}. 

Interestingly, two phonon modes with wave numbers 197 cm$^{-1}$ ($\sim$ 6.0 THz) and
339 cm$^{-1}$ ($\sim$ 10.2 THz) show the greatest projection amplitudes. 
Polarization vectors of these two phonon
modes are also shown in Figs.~\ref{fig:VIB}(e) and \ref{fig:VIB}(f), and the 
vibration patterns of these two phonon modes also match nicely with the observed first two stages 
of the structural evolution after photoexcitations~\cite{Baum02112007}.

Combining this result with the analysis of forces exerted on individual atoms 
upon ``optical doping" of electrons and holes, we have clearly identified the active phonon modes
involved in the photo-induced insulator-metal transition of VO$_2$. In addition, our result 
indicates that the strong coupling between the 339 cm$^{-1}$ (10.2 THz) phonon and hole states
is responsible for the rapid V-V separation upon photoexcitation, whereas the coupling between
the 197 cm$^{-1}$ ($\sim$ 6.0 THz) phonon and electrons explains the zag-zig motion of V atoms 
at a longer timescale. The 6.0 THz phonon mode is often observed in coherent phonon
experiment. However, the V-V separation happens at a much faster timescale~\cite{Baum02112007} so that
the corresponding phonon mode (10.2 THz) cannot be observed in coherent phonon experiment.
In addition to these two phonon modes, there are several other modes
that have large projection amplitudes. In particular, the 224 cm$^{-1}$ ($\sim$ 6.72 THz)
phonon mode may be related to the 6.75 THz phonon modes observed in coherent phonon
experiments~\cite{PhysRevB.70.161102}.

There are still puzzles that need to be resolved before we can fully untangle the experimental
findings. As mentioned above, it is very likely that the ultrafast ($\sim 100$ fs) 
photo-induced insulator-metal transition and the rapid V-V separation upon photoexcitation 
are closely related, and uncovering their cause-effect relationship 
is the last step toward a full understanding of the physics of the
phase transition in VO$_2$. So far our results are consistent with experiments,
but we have not addressed the issue regarding the possible connection between 
the observed rapid V-V separation and the ultrafast photo-induced insulator-metal transition.

\begin{figure}[!t]
\includegraphics[width=7cm]{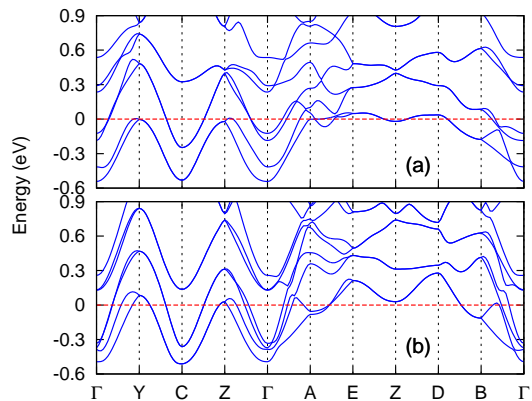}
\caption{\label{fig:M1esmov} (Color online) 
(a) Calculated band structure using a distorted M$_1$ structure in which atoms are displaced 
according to the projection of the atomic displacement from the M$_1$ to the R phase 
onto the 339 cm$^{-1}$ phonon mode. (b) Calculated band structure using the 
crystal structure relaxed with 0.15 holes per VO$_2$ formula. 
The Fermi level is indicated by the dashed horizontal line at E=0.}
\end{figure}

To this end, we have carried out two additional calculations. In the first calculation, 
the ideal M$_1$ crystal structure is displaced according to the projection
of the displacement vector $\Delta\mathbf{R}$ in Eq.~(\ref{eq:eq1}) onto the 339 cm$^{-1}$ (10.2 THz)
phonon mode. This procedure naturally results in a separation of V-V dimers. A band structure
calculation is then carried out using the distorted structure. 
In the second calculation, we introduce 0.15 holes per VO$_2$ formula 
unit into the system. The internal coordinates are then relaxed while the lattice
vectors are fixed. A band structure calculation
is then carried out using the relaxed structure with the presence of holes.
Since photo-induced ultrafast insulator-metal transition occurs long before a full
conversion of the lattice geometry from monoclinic to rutile, our calculations are carried out
using the lattice vectors of the M$_1$ phase. The resulting band structures are shown in Fig.~\ref{fig:M1esmov}. 
Both band structures clearly show a metallic behavior beyond what is expected for a simple semiconductor
with small structural distortions: There is a massive reorganization of the
electronic states are the V-V bonds break.
These results clearly illustrate the much discussed strong coupling between 
the lattice and electronic degrees of freedom in this system. However, not all
lattice degrees of freedom are coupled equally with electrons. It is the phonon mode
involving the separation of V-V dimer that has the most important 
impact on the underlying electronic structure. It is the rapid separation of
V-V pairs (i.e., breaking of the V-V dimers), which itself is a direct result 
of a strong phonon-hole coupling, that is responsible for the observed 
ultrafast photo-induced insulator-metal transition. 

The process of ultrafast photo-induced insulator-metal transition in VO$_2$ can now be summarized as follows: 
photoexcitation results in a depletion of electrons in the bonding states which
are critical for V-V dimerization and the insulating behavior of the M$_1$ phase. When this weakening of the
V-V bonds reaches a critical point (under critical laser fluence), the V-V bond breaks.
This bond breaking (V-V separation) is a result of a strong lattice-hole coupling
in this system. The breaking of the V-V bonds in turn leads to an instantaneous 
electronic phase transition from the insulating
to a metallic state. Note that at this stage, the lattice remains monoclinic in shape.
Electron-lattice coupling and thermalization eventually eliminate the zig-zag 
arrangement of the V atoms and fully convert both the internal structure and the shape 
of the crystal from the M$_1$ to the R phase. The corresponding subnanosecond lattice dynamics
is associated with the 6.0 THz phonon often observed in coherent phonon experiment.

%\begin{acknowledgements}
This work is supported in part by NSFC under grant numbers 
11234012 and 50821064, by Shanghai Key Basic Research Project (3109DJ1400201), 
and by the CAS/SAFEA International Partnership Program for Creative Research Teams.
PZ is supported by the US National Science Foundation
under Grant No. DMR-0946404 and by the US Department
of Energy under Grant No. DE-SC0002623.
%\end{acknowledgements}

\bibliography{reference}

\begin{thebibliography}{17}%
\makeatletter
\providecommand \@ifxundefined [1]{%
 \@ifx{#1\undefined}
}%
\providecommand \@ifnum [1]{%
 \ifnum #1\expandafter \@firstoftwo
 \else \expandafter \@secondoftwo
 \fi
}%
\providecommand \@ifx [1]{%
 \ifx #1\expandafter \@firstoftwo
 \else \expandafter \@secondoftwo
 \fi
}%
\providecommand \natexlab [1]{#1}%
\providecommand \enquote  [1]{``#1''}%
\providecommand \bibnamefont  [1]{#1}%
\providecommand \bibfnamefont [1]{#1}%
\providecommand \citenamefont [1]{#1}%
\providecommand \href@noop [0]{\@secondoftwo}%
\providecommand \href [0]{\begingroup \@sanitize@url \@href}%
\providecommand \@href[1]{\@@startlink{#1}\@@href}%
\providecommand \@@href[1]{\endgroup#1\@@endlink}%
\providecommand \@sanitize@url [0]{\catcode `\\12\catcode `\$12\catcode
  `\&12\catcode `\#12\catcode `\^12\catcode `\_12\catcode `\%12\relax}%
\providecommand \@@startlink[1]{}%
\providecommand \@@endlink[0]{}%
\providecommand \url  [0]{\begingroup\@sanitize@url \@url }%
\providecommand \@url [1]{\endgroup\@href {#1}{\urlprefix }}%
\providecommand \urlprefix  [0]{URL }%
\providecommand \Eprint [0]{\href }%
\providecommand \doibase [0]{http://dx.doi.org/}%
\providecommand \selectlanguage [0]{\@gobble}%
\providecommand \bibinfo  [0]{\@secondoftwo}%
\providecommand \bibfield  [0]{\@secondoftwo}%
\providecommand \translation [1]{[#1]}%
\providecommand \BibitemOpen [0]{}%
\providecommand \bibitemStop [0]{}%
\providecommand \bibitemNoStop [0]{.\EOS\space}%
\providecommand \EOS [0]{\spacefactor3000\relax}%
\providecommand \BibitemShut  [1]{\csname bibitem#1\endcsname}%
\let\auto@bib@innerbib\@empty
%</preamble>
\bibitem [{\citenamefont {Morin}(1959)}]{PhysRevLett.3.34}%
  \BibitemOpen
  \bibfield  {author} {\bibinfo {author} {\bibfnamefont {F.~J.}\ \bibnamefont
  {Morin}},\ }\href {\doibase 10.1103/PhysRevLett.3.34} {\bibfield  {journal}
  {\bibinfo  {journal} {Phys. Rev. Lett.}\ }\textbf {\bibinfo {volume} {3}},\
  \bibinfo {pages} {34} (\bibinfo {year} {1959})}\BibitemShut {NoStop}%
\bibitem [{\citenamefont {Cavalleri}\ \emph {et~al.}(2001)\citenamefont
  {Cavalleri}, \citenamefont {T\'oth}, \citenamefont {Siders}, \citenamefont
  {Squier}, \citenamefont {R\'aksi}, \citenamefont {Forget},\ and\
  \citenamefont {Kieffer}}]{PhysRevLett.87.237401}%
  \BibitemOpen
  \bibfield  {author} {\bibinfo {author} {\bibfnamefont {A.}~\bibnamefont
  {Cavalleri}}, \bibinfo {author} {\bibfnamefont {C.}~\bibnamefont {T\'oth}},
  \bibinfo {author} {\bibfnamefont {C.~W.}\ \bibnamefont {Siders}}, \bibinfo
  {author} {\bibfnamefont {J.~A.}\ \bibnamefont {Squier}}, \bibinfo {author}
  {\bibfnamefont {F.}~\bibnamefont {R\'aksi}}, \bibinfo {author} {\bibfnamefont
  {P.}~\bibnamefont {Forget}}, \ and\ \bibinfo {author} {\bibfnamefont {J.~C.}\
  \bibnamefont {Kieffer}},\ }\href {\doibase 10.1103/PhysRevLett.87.237401}
  {\bibfield  {journal} {\bibinfo  {journal} {Phys. Rev. Lett.}\ }\textbf
  {\bibinfo {volume} {87}},\ \bibinfo {pages} {237401} (\bibinfo {year}
  {2001})}\BibitemShut {NoStop}%
\bibitem [{\citenamefont {Cavalleri}\ \emph
  {et~al.}(2004{\natexlab{a}})\citenamefont {Cavalleri}, \citenamefont {Chong},
  \citenamefont {Fourmaux}, \citenamefont {Glover}, \citenamefont {Heimann},
  \citenamefont {Kieffer}, \citenamefont {Mun}, \citenamefont {Padmore},\ and\
  \citenamefont {Schoenlein}}]{PhysRevB.69.153106}%
  \BibitemOpen
  \bibfield  {author} {\bibinfo {author} {\bibfnamefont {A.}~\bibnamefont
  {Cavalleri}}, \bibinfo {author} {\bibfnamefont {H.~H.~W.}\ \bibnamefont
  {Chong}}, \bibinfo {author} {\bibfnamefont {S.}~\bibnamefont {Fourmaux}},
  \bibinfo {author} {\bibfnamefont {T.~E.}\ \bibnamefont {Glover}}, \bibinfo
  {author} {\bibfnamefont {P.~A.}\ \bibnamefont {Heimann}}, \bibinfo {author}
  {\bibfnamefont {J.~C.}\ \bibnamefont {Kieffer}}, \bibinfo {author}
  {\bibfnamefont {B.~S.}\ \bibnamefont {Mun}}, \bibinfo {author} {\bibfnamefont
  {H.~A.}\ \bibnamefont {Padmore}}, \ and\ \bibinfo {author} {\bibfnamefont
  {R.~W.}\ \bibnamefont {Schoenlein}},\ }\href {\doibase
  10.1103/PhysRevB.69.153106} {\bibfield  {journal} {\bibinfo  {journal} {Phys.
  Rev. B}\ }\textbf {\bibinfo {volume} {69}},\ \bibinfo {pages} {153106}
  (\bibinfo {year} {2004}{\natexlab{a}})}\BibitemShut {NoStop}%
\bibitem [{\citenamefont {Cavalleri}\ \emph
  {et~al.}(2004{\natexlab{b}})\citenamefont {Cavalleri}, \citenamefont
  {Dekorsy}, \citenamefont {Chong}, \citenamefont {Kieffer},\ and\
  \citenamefont {Schoenlein}}]{PhysRevB.70.161102}%
  \BibitemOpen
  \bibfield  {author} {\bibinfo {author} {\bibfnamefont {A.}~\bibnamefont
  {Cavalleri}}, \bibinfo {author} {\bibfnamefont {T.}~\bibnamefont {Dekorsy}},
  \bibinfo {author} {\bibfnamefont {H.~H.~W.}\ \bibnamefont {Chong}}, \bibinfo
  {author} {\bibfnamefont {J.~C.}\ \bibnamefont {Kieffer}}, \ and\ \bibinfo
  {author} {\bibfnamefont {R.~W.}\ \bibnamefont {Schoenlein}},\ }\href
  {\doibase 10.1103/PhysRevB.70.161102} {\bibfield  {journal} {\bibinfo
  {journal} {Phys. Rev. B}\ }\textbf {\bibinfo {volume} {70}},\ \bibinfo
  {pages} {161102} (\bibinfo {year} {2004}{\natexlab{b}})}\BibitemShut
  {NoStop}%
\bibitem [{\citenamefont {K\"ubler}\ \emph {et~al.}(2007)\citenamefont
  {K\"ubler}, \citenamefont {Ehrke}, \citenamefont {Huber}, \citenamefont
  {Lopez}, \citenamefont {Halabica}, \citenamefont {Haglund},\ and\
  \citenamefont {Leitenstorfer}}]{PhysRevLett.99.116401}%
  \BibitemOpen
  \bibfield  {author} {\bibinfo {author} {\bibfnamefont {C.}~\bibnamefont
  {K\"ubler}}, \bibinfo {author} {\bibfnamefont {H.}~\bibnamefont {Ehrke}},
  \bibinfo {author} {\bibfnamefont {R.}~\bibnamefont {Huber}}, \bibinfo
  {author} {\bibfnamefont {R.}~\bibnamefont {Lopez}}, \bibinfo {author}
  {\bibfnamefont {A.}~\bibnamefont {Halabica}}, \bibinfo {author}
  {\bibfnamefont {R.~F.}\ \bibnamefont {Haglund}}, \ and\ \bibinfo {author}
  {\bibfnamefont {A.}~\bibnamefont {Leitenstorfer}},\ }\href {\doibase
  10.1103/PhysRevLett.99.116401} {\bibfield  {journal} {\bibinfo  {journal}
  {Phys. Rev. Lett.}\ }\textbf {\bibinfo {volume} {99}},\ \bibinfo {pages}
  {116401} (\bibinfo {year} {2007})}\BibitemShut {NoStop}%
\bibitem [{\citenamefont {Lysenko}\ \emph {et~al.}(2007)\citenamefont
  {Lysenko}, \citenamefont {R\'ua}, \citenamefont {Vikhnin}, \citenamefont
  {Fern\'andez},\ and\ \citenamefont {Liu}}]{PhysRevB.76.035104}%
  \BibitemOpen
  \bibfield  {author} {\bibinfo {author} {\bibfnamefont {S.}~\bibnamefont
  {Lysenko}}, \bibinfo {author} {\bibfnamefont {A.}~\bibnamefont {R\'ua}},
  \bibinfo {author} {\bibfnamefont {V.}~\bibnamefont {Vikhnin}}, \bibinfo
  {author} {\bibfnamefont {F.}~\bibnamefont {Fern\'andez}}, \ and\ \bibinfo
  {author} {\bibfnamefont {H.}~\bibnamefont {Liu}},\ }\href {\doibase
  10.1103/PhysRevB.76.035104} {\bibfield  {journal} {\bibinfo  {journal} {Phys.
  Rev. B}\ }\textbf {\bibinfo {volume} {76}},\ \bibinfo {pages} {035104}
  (\bibinfo {year} {2007})}\BibitemShut {NoStop}%
\bibitem [{\citenamefont {Nakajima}\ \emph {et~al.}(2008)\citenamefont
  {Nakajima}, \citenamefont {Takubo}, \citenamefont {Hiroi}, \citenamefont
  {Ueda},\ and\ \citenamefont {Suemoto}}]{nakajima011907}%
  \BibitemOpen
  \bibfield  {author} {\bibinfo {author} {\bibfnamefont {M.}~\bibnamefont
  {Nakajima}}, \bibinfo {author} {\bibfnamefont {N.}~\bibnamefont {Takubo}},
  \bibinfo {author} {\bibfnamefont {Z.}~\bibnamefont {Hiroi}}, \bibinfo
  {author} {\bibfnamefont {Y.}~\bibnamefont {Ueda}}, \ and\ \bibinfo {author}
  {\bibfnamefont {T.}~\bibnamefont {Suemoto}},\ }\href {\doibase
  10.1063/1.2830664} {\bibfield  {journal} {\bibinfo  {journal} {Appl. Phys.
  Lett.}\ }\textbf {\bibinfo {volume} {92}},\ \bibinfo {eid} {011907} (\bibinfo
  {year} {2008})}\BibitemShut {NoStop}%
\bibitem [{\citenamefont {Pashkin}\ \emph {et~al.}(2011)\citenamefont
  {Pashkin}, \citenamefont {K\"ubler}, \citenamefont {Ehrke}, \citenamefont
  {Lopez}, \citenamefont {Halabica}, \citenamefont {Haglund}, \citenamefont
  {Huber},\ and\ \citenamefont {Leitenstorfer}}]{PhysRevB.83.195120}%
  \BibitemOpen
  \bibfield  {author} {\bibinfo {author} {\bibfnamefont {A.}~\bibnamefont
  {Pashkin}}, \bibinfo {author} {\bibfnamefont {C.}~\bibnamefont {K\"ubler}},
  \bibinfo {author} {\bibfnamefont {H.}~\bibnamefont {Ehrke}}, \bibinfo
  {author} {\bibfnamefont {R.}~\bibnamefont {Lopez}}, \bibinfo {author}
  {\bibfnamefont {A.}~\bibnamefont {Halabica}}, \bibinfo {author}
  {\bibfnamefont {R.~F.}\ \bibnamefont {Haglund}}, \bibinfo {author}
  {\bibfnamefont {R.}~\bibnamefont {Huber}}, \ and\ \bibinfo {author}
  {\bibfnamefont {A.}~\bibnamefont {Leitenstorfer}},\ }\href {\doibase
  10.1103/PhysRevB.83.195120} {\bibfield  {journal} {\bibinfo  {journal} {Phys.
  Rev. B}\ }\textbf {\bibinfo {volume} {83}},\ \bibinfo {pages} {195120}
  (\bibinfo {year} {2011})}\BibitemShut {NoStop}%
\bibitem [{\citenamefont {Cocker}\ \emph {et~al.}(2012)\citenamefont {Cocker},
  \citenamefont {Titova}, \citenamefont {Fourmaux}, \citenamefont {Holloway},
  \citenamefont {Bandulet}, \citenamefont {Brassard}, \citenamefont {Kieffer},
  \citenamefont {El~Khakani},\ and\ \citenamefont
  {Hegmann}}]{PhysRevB.85.155120}%
  \BibitemOpen
  \bibfield  {author} {\bibinfo {author} {\bibfnamefont {T.~L.}\ \bibnamefont
  {Cocker}}, \bibinfo {author} {\bibfnamefont {L.~V.}\ \bibnamefont {Titova}},
  \bibinfo {author} {\bibfnamefont {S.}~\bibnamefont {Fourmaux}}, \bibinfo
  {author} {\bibfnamefont {G.}~\bibnamefont {Holloway}}, \bibinfo {author}
  {\bibfnamefont {H.-C.}\ \bibnamefont {Bandulet}}, \bibinfo {author}
  {\bibfnamefont {D.}~\bibnamefont {Brassard}}, \bibinfo {author}
  {\bibfnamefont {J.-C.}\ \bibnamefont {Kieffer}}, \bibinfo {author}
  {\bibfnamefont {M.~A.}\ \bibnamefont {El~Khakani}}, \ and\ \bibinfo {author}
  {\bibfnamefont {F.~A.}\ \bibnamefont {Hegmann}},\ }\href {\doibase
  10.1103/PhysRevB.85.155120} {\bibfield  {journal} {\bibinfo  {journal} {Phys.
  Rev. B}\ }\textbf {\bibinfo {volume} {85}},\ \bibinfo {pages} {155120}
  (\bibinfo {year} {2012})}\BibitemShut {NoStop}%
\bibitem [{\citenamefont {Baum}\ \emph {et~al.}(2007)\citenamefont {Baum},
  \citenamefont {Yang},\ and\ \citenamefont {Zewail}}]{Baum02112007}%
  \BibitemOpen
  \bibfield  {author} {\bibinfo {author} {\bibfnamefont {P.}~\bibnamefont
  {Baum}}, \bibinfo {author} {\bibfnamefont {D.-S.}\ \bibnamefont {Yang}}, \
  and\ \bibinfo {author} {\bibfnamefont {A.~H.}\ \bibnamefont {Zewail}},\
  }\href {\doibase 10.1126/science.1147724} {\bibfield  {journal} {\bibinfo
  {journal} {Science}\ }\textbf {\bibinfo {volume} {318}},\ \bibinfo {pages}
  {788} (\bibinfo {year} {2007})}\BibitemShut {NoStop}%
\bibitem [{\citenamefont {Kim}\ \emph {et~al.}(2006)\citenamefont {Kim},
  \citenamefont {Lee}, \citenamefont {Kim}, \citenamefont {Chae}, \citenamefont
  {Yun}, \citenamefont {Kang}, \citenamefont {Han}, \citenamefont {Yee},\ and\
  \citenamefont {Lim}}]{PhysRevLett.97.266401}%
  \BibitemOpen
  \bibfield  {author} {\bibinfo {author} {\bibfnamefont {H.-T.}\ \bibnamefont
  {Kim}}, \bibinfo {author} {\bibfnamefont {Y.~W.}\ \bibnamefont {Lee}},
  \bibinfo {author} {\bibfnamefont {B.-J.}\ \bibnamefont {Kim}}, \bibinfo
  {author} {\bibfnamefont {B.-G.}\ \bibnamefont {Chae}}, \bibinfo {author}
  {\bibfnamefont {S.~J.}\ \bibnamefont {Yun}}, \bibinfo {author} {\bibfnamefont
  {K.-Y.}\ \bibnamefont {Kang}}, \bibinfo {author} {\bibfnamefont {K.-J.}\
  \bibnamefont {Han}}, \bibinfo {author} {\bibfnamefont {K.-J.}\ \bibnamefont
  {Yee}}, \ and\ \bibinfo {author} {\bibfnamefont {Y.-S.}\ \bibnamefont
  {Lim}},\ }\href {\doibase 10.1103/PhysRevLett.97.266401} {\bibfield
  {journal} {\bibinfo  {journal} {Phys. Rev. Lett.}\ }\textbf {\bibinfo
  {volume} {97}},\ \bibinfo {pages} {266401} (\bibinfo {year}
  {2006})}\BibitemShut {NoStop}%
\bibitem [{\citenamefont {Tao}\ \emph {et~al.}(2012)\citenamefont {Tao},
  \citenamefont {Han}, \citenamefont {Mahanti}, \citenamefont {Duxbury},
  \citenamefont {Yuan}, \citenamefont {Ruan}, \citenamefont {Wang},\ and\
  \citenamefont {Wu}}]{PhysRevLett.109.166406}%
  \BibitemOpen
  \bibfield  {author} {\bibinfo {author} {\bibfnamefont {Z.}~\bibnamefont
  {Tao}}, \bibinfo {author} {\bibfnamefont {T.-R.~T.}\ \bibnamefont {Han}},
  \bibinfo {author} {\bibfnamefont {S.~D.}\ \bibnamefont {Mahanti}}, \bibinfo
  {author} {\bibfnamefont {P.~M.}\ \bibnamefont {Duxbury}}, \bibinfo {author}
  {\bibfnamefont {F.}~\bibnamefont {Yuan}}, \bibinfo {author} {\bibfnamefont
  {C.-Y.}\ \bibnamefont {Ruan}}, \bibinfo {author} {\bibfnamefont
  {K.}~\bibnamefont {Wang}}, \ and\ \bibinfo {author} {\bibfnamefont
  {J.}~\bibnamefont {Wu}},\ }\href {\doibase 10.1103/PhysRevLett.109.166406}
  {\bibfield  {journal} {\bibinfo  {journal} {Phys. Rev. Lett.}\ }\textbf
  {\bibinfo {volume} {109}},\ \bibinfo {pages} {166406} (\bibinfo {year}
  {2012})}\BibitemShut {NoStop}%
\bibitem [{\citenamefont {Longo}\ and\ \citenamefont
  {Kierkegaard}(1970)}]{M1struc}%
  \BibitemOpen
  \bibfield  {author} {\bibinfo {author} {\bibfnamefont {J.~M.}\ \bibnamefont
  {Longo}}\ and\ \bibinfo {author} {\bibfnamefont {P.}~\bibnamefont
  {Kierkegaard}},\ }\href {\doibase 10.3891/acta.chem.scand.24-0420} {\bibfield
   {journal} {\bibinfo  {journal} {Acta Chim. Scand.}\ }\textbf {\bibinfo
  {volume} {24}},\ \bibinfo {pages} {420} (\bibinfo {year} {1970})}\BibitemShut
  {NoStop}%
\bibitem [{\citenamefont {Bl\"ochl}(1994)}]{PhysRevB.50.17953}%
  \BibitemOpen
  \bibfield  {author} {\bibinfo {author} {\bibfnamefont {P.~E.}\ \bibnamefont
  {Bl\"ochl}},\ }\href {\doibase 10.1103/PhysRevB.50.17953} {\bibfield
  {journal} {\bibinfo  {journal} {Phys. Rev. B}\ }\textbf {\bibinfo {volume}
  {50}},\ \bibinfo {pages} {17953} (\bibinfo {year} {1994})}\BibitemShut
  {NoStop}%
\bibitem [{\citenamefont {Kresse}\ and\ \citenamefont
  {Furthm\"uller}(1996)}]{PhysRevB.54.11169}%
  \BibitemOpen
  \bibfield  {author} {\bibinfo {author} {\bibfnamefont {G.}~\bibnamefont
  {Kresse}}\ and\ \bibinfo {author} {\bibfnamefont {J.}~\bibnamefont
  {Furthm\"uller}},\ }\href {\doibase 10.1103/PhysRevB.54.11169} {\bibfield
  {journal} {\bibinfo  {journal} {Phys. Rev. B}\ }\textbf {\bibinfo {volume}
  {54}},\ \bibinfo {pages} {11169} (\bibinfo {year} {1996})}\BibitemShut
  {NoStop}%
\bibitem [{\citenamefont {Schilbe}\ and\ \citenamefont
  {Maurer}(2004)}]{Schilbe2004449}%
  \BibitemOpen
  \bibfield  {author} {\bibinfo {author} {\bibfnamefont {P.}~\bibnamefont
  {Schilbe}}\ and\ \bibinfo {author} {\bibfnamefont {D.}~\bibnamefont
  {Maurer}},\ }\href {\doibase 10.1016/j.msea.2003.08.114} {\bibfield
  {journal} {\bibinfo  {journal} {Mater. Sci. Eng., A}\ }\textbf {\bibinfo
  {volume} {370}},\ \bibinfo {pages} {449 } (\bibinfo {year}
  {2004})}\BibitemShut {NoStop}%
\bibitem [{\citenamefont {Barker}\ \emph {et~al.}(1966)\citenamefont {Barker},
  \citenamefont {Verleur},\ and\ \citenamefont
  {Guggenheim}}]{PhysRevLett.17.1286}%
  \BibitemOpen
  \bibfield  {author} {\bibinfo {author} {\bibfnamefont {A.~S.}\ \bibnamefont
  {Barker}}, \bibinfo {author} {\bibfnamefont {H.~W.}\ \bibnamefont {Verleur}},
  \ and\ \bibinfo {author} {\bibfnamefont {H.~J.}\ \bibnamefont {Guggenheim}},\
  }\href {\doibase 10.1103/PhysRevLett.17.1286} {\bibfield  {journal} {\bibinfo
   {journal} {Phys. Rev. Lett.}\ }\textbf {\bibinfo {volume} {17}},\ \bibinfo
  {pages} {1286} (\bibinfo {year} {1966})}\BibitemShut {NoStop}%
\end{thebibliography}%

\end{document}